# Strong Transient Flows Generated by Thermoplasmonic Bubble Nucleation

Steven Jones[1], Daniel Andrén[1], Tomasz J. Antosiewicz[2], Alexander Stilgoe[3], Halina Rubinsztein-Dunlop[3], & Mikael Käll[1]

1.  Department of Physics, Chalmers University of Technology, 41296, Gothenburg, Sweden.
2.  Faculty of Physics, University of Warsaw, 02-093, Warsaw, Poland.
3.  School of Mathematics and Physics, University of Queensland, 4072, Queensland, Australia.

**ABSTRACT:** The challenge of inducing and controlling localized fluid flows for generic force actuation and for achieving efficient mass transport in microfluidics is key to the development of next generation miniaturized systems for chemistry and life sciences. Here we demonstrate a methodology for the robust generation and precise quantification of extremely strong flow transients driven by vapor bubble nucleation on spatially isolated plasmonic nanoantennas excited by light. The system is capable of producing peak flow speeds of the order mm/s at modulation rates up to ≈100 Hz in water, thus allowing for a variety of high-throughput applications. Analysis of flow dynamics and fluid viscosity dependence indicates that the transient originates in the rapid bubble expansion that follows nucleation rather than being strictly thermocapillary in nature.

Achieving efficient liquid transport and mixing at the microscale in spite of the obstacles posed by low Reynolds numbers and laminar flow is integral to microfluidic system development. For instance, detection of analytes at very low concentration remains a key challenge in surface-based biosensing since signal integration times can easily be limited by analyte diffusion.[1,2] Thus, active transport is typically needed to increase the detection rate so as to reach reasonable assay times. In other applications, e.g. in the development of micromachines and miniaturized engines or motors, the ability to apply large forces mediated by strong localized flows in confined spaces is crucial to or may extend the functionality of such devices.[3–5]

Photothermally induced surface microbubbles show great promise as facile microscopic force and flow actuators since optical control can be achieved through contactless external illumination without the need for any mechanical or electrical interconnects. Bubble generation is typically based on plasmon-enhanced absorption of continuous-wave laser light in semi-continuous metal films to locally superheat water and induce bubble nucleation.[6,7] The optically established temperature profile around the bubble generates fluid flow via the thermocapillary effect,[8] resulting in a flow profile well approximated by a Stokeslet near an interface,[9] and can impart forces over a large volume.[10] The ability to generate localized fluid flow near interfaces is particularly impactful in microfluidics as this is the region where externally driven flow is least effective at mass transport due to the no slip condition at the boundary (e.g. Poiseuille flows).

Figure 1a schematically illustrates the current understanding of thermoplasmonic bubble formation and dissipation in water.[6,7,11,12] The onset of light absorption in the metal near-instantaneously raises its surface-temperature and, for moderate laser powers, causes significant superheating in the surrounding liquid. Beyond a critical local temperature threshold, typically well above the macroscopic boiling point of water,[13] a bubble can be nucleated via vaporization. Since the surrounding hot fluid is now locally oversaturated, there will be an influx of gas into the bubble causing further growth. This growth stage can continue nearly unimpeded if the heat source is large. After the heating laser is turned off the bubble immediately contracts due to a drop in internal pressure and vapour-condensation. Further bubble dissipation proceeds via expulsion of gas back into the surrounding fluid, a process that is diffusion limited and therefore can be extremely slow for large bubbles.[14,15]

Recent results on thermoplasmonic bubbles has demonstrated their potential for improving sensing via analyte accumulation,[16,17] for physical deposition[18–20] and for chemical synthesis.[21,22] However, previous works in air-equilibrated water often involve large bubbles that can exhibit semi-continuous growth and require hundreds to thousands of seconds to fully dissipate.[6] In contrast, for many applications it would be highly advantageous to instead generate small bubbles able to produce strong flows at high modulation rates in order to be utilized in very confined environments and in high throughput systems. The key to such rapid dynamic control of the bubble life cycle is to prohibit the formation of large bubbles to ensure that the gaseous growth stage is quickly terminated.[23] The transient behaviour of bubble formation and dissipation, and the resulting flow response, now becomes critical for overall system design.

Here we demonstrate that by utilizing isolated plasmonic nanoantennas for heat generation, it is possible to generate small and quickly dissipating thermoplasmonic bubbles in air-equilibrated water. These bubbles are able to produce strong flow transients with flow velocities >1 mm/s several microns away from the bubble at modulation rates up to ≈100 Hz in water, orders of magnitude faster than is possible with larger microbubbles. The bubbles are observed to generate localized Stokeslet-like flow patterns well suited for analyte accumulation and the persistent flow following the transient are found to be comparable in magnitude (on the order of 100 μm/s) to those generated previously with larger bubbles. The transient behaviour is found to occur from a different driving mechanism (mechanical) than the persistent flow component (thermocapillary). Its dominant character with regard to peak flow-speed provides an avenue for further optimization as a highly advantageous force actuator for integration in next generation microfluidic systems.

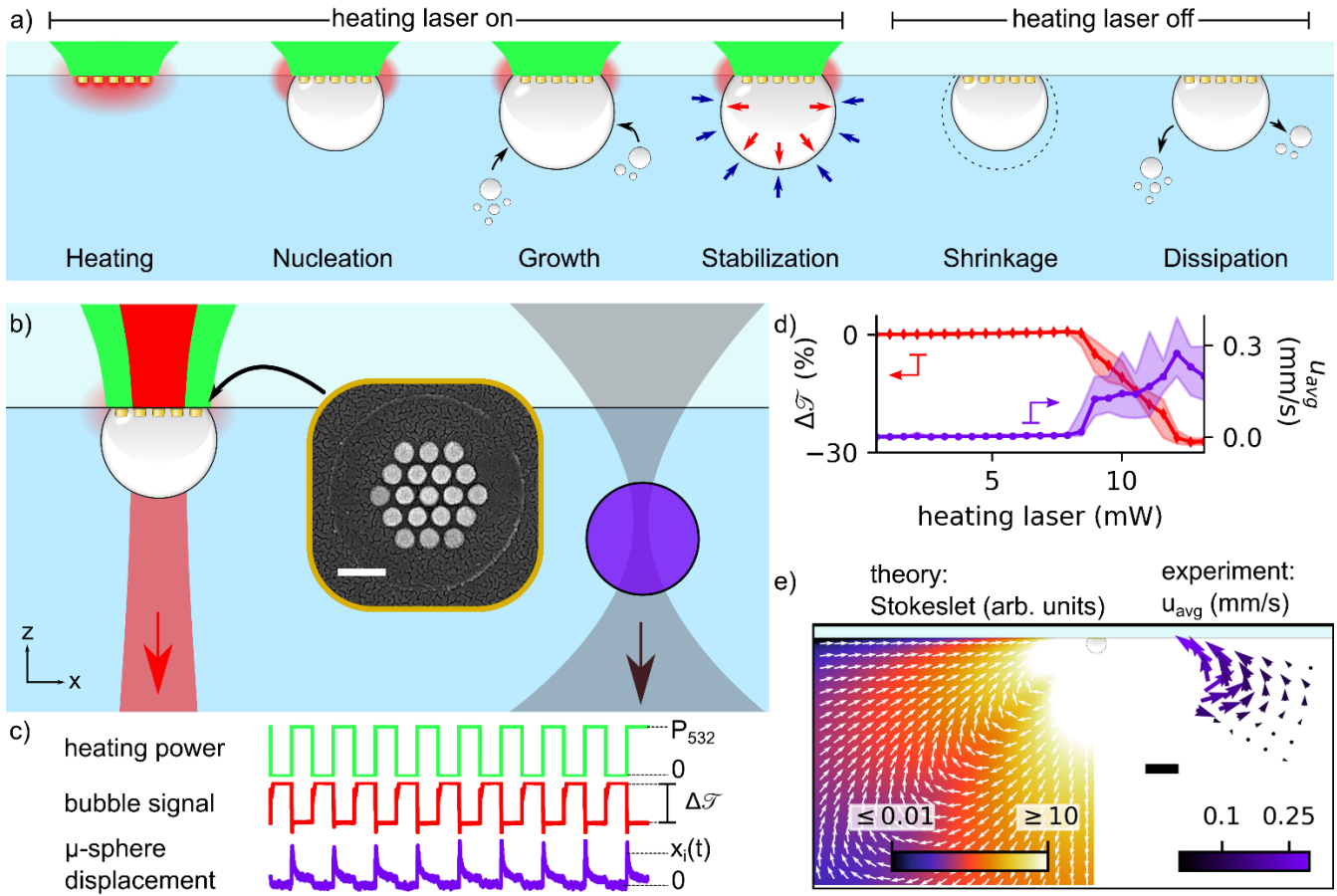

*Figure 1. Making bubbles and measuring flows. a) Overview of the bubble life cycle during heat modulation. b) Schematic of the experimental environment. b-inset) Scanning electron micrograph of a nanoantenna (after being used for bubble generation). Scale bar is 300 nm. c) Experimental procedure showing heating laser modulation (top), changes in detection laser transmission due to bubble formation/dissipation (middle), and probe particle displacement due to bubble induced flow (bottom). d) Changes in detection laser transmission (left y-axis) and local flow (right y-axis) for a range of heating laser powers. Data show the average transmission (flow speed) during the on state. Measurement location: $x_0$, $z_0$ = 7, -1.5 μm. Points (shaded region) show the mean (range) from 8 measurements on 7 antennas. e) Theoretical (left) and experimental (right) flow speed profile in the x-z plane due to the presence of a photothermally generated bubble. Scale bar is 2 μm. Note that the maximum Reynolds number for the experimental data in e) is approximately $10^{-3}$.*

**Results & Discussion**

Bubble generation occurs on an isolated nanoantenna (Figure 1b-inset) submersed in air-equilibrated water. The thermoplasmonic antenna is heated with a continuous wave 532 nm laser beam in resonance with the d-band transitions of Au and the localized dipolar plasmon modes of the structure. This thermoplasmonic antenna configuration was chosen so that the optical and thermal characteristics could be independently controlled by adjusting the size of the individual disk and overall configuration, respectively (further details are given in Methods - Thermoplasmonic Antenna: Design & Fabrication). The thermoplasmonic structure utilized here facilitates the generation of small bubbles in air-equilibrated water where bubble growth is quickly terminated by the Laplace pressure and limited volume of heated water. This maintains the bubble radius at approximately $r \approx 500$ nm for the structures used here (depending on heating laser power[23]) which enables quick dissipation, after the heating laser is turned off, and therefore high modulation rates. Importantly, the dissipation time for these bubbles, and therefore maximum modulation rate, depends only on the thermoplasmonic antenna size and heating laser power and is not significantly influenced by heating duration.

We use optical force microscopy[24] in conjunction with optical bubble generation and detection in order to study the transient fluid response during the bubble modulation cycle. Figure 1b provides an overview of the measurement procedure (see supporting information SI-1 and the methods section for details of the experimental setup). The heating laser is modulated in a square wave temporal profile (duty-cycle = 50%) at a pre-set frequency $f_{mod}$ and power $P_{532}$ (Figure 1c, upper trace). The bubble is monitored by changes in transmission $\Delta T_{633}$ of a low power 633 nm laser beam coaligned with the heating laser since the bubbles are too small and quickly dissipating to be accurately detected with standard brightfield microscopy (Figure 1c, middle trace). Note that $\Delta T_{633}$ exhibits a complex variation with bubble radius and contact angle and the signal can therefore not be directly translated to bubble size (see supporting information S2b).

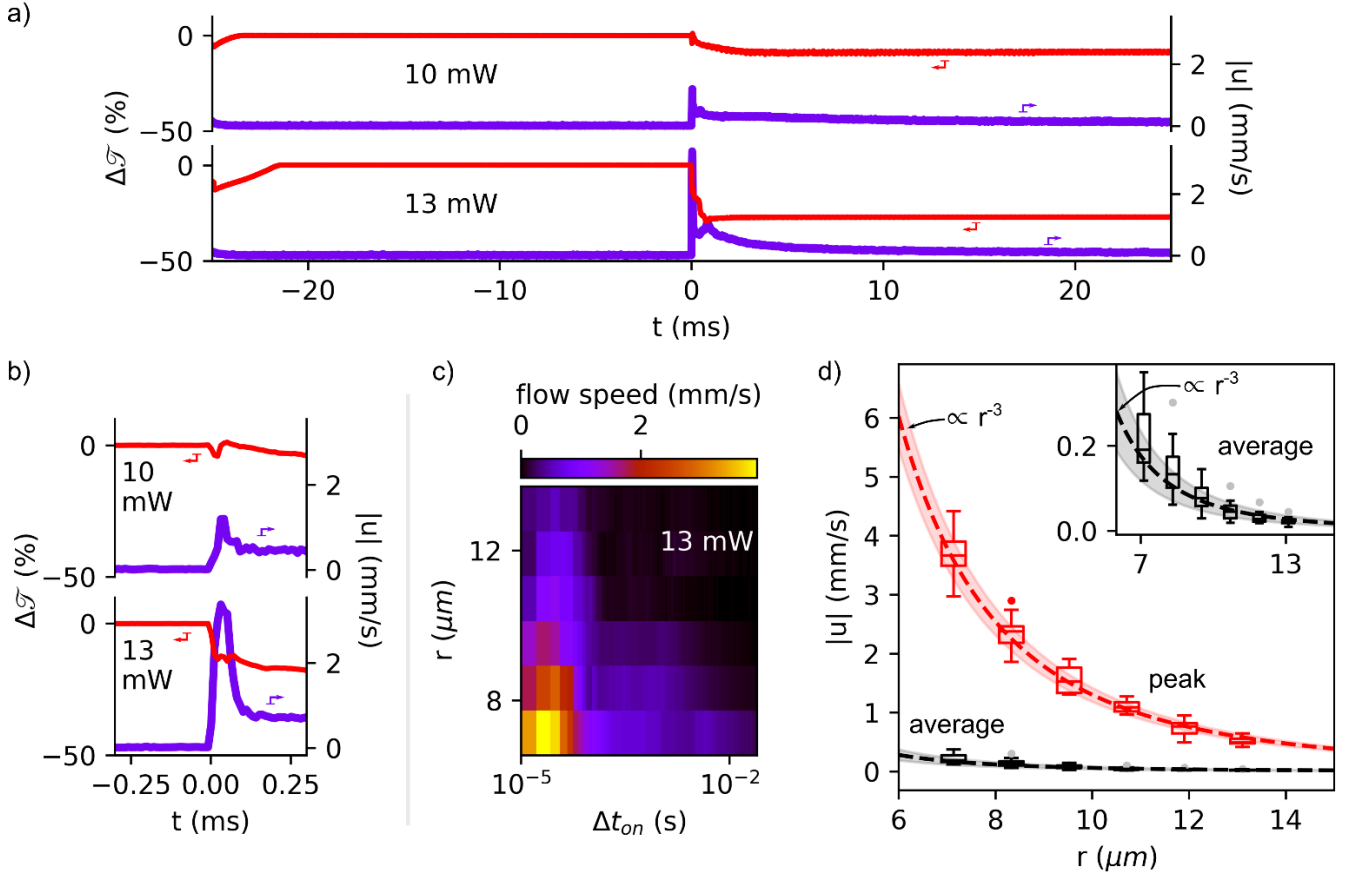

*Figure 2. Transient flow dynamics. a) Full cycle (20 Hz) trace of the bubble induced transmission ($\Delta T_{633}$, left y-axis) and magnitude of local flow ($|u|$, right y-axis) measured at $x_0, z_0 = 7, -1.5$ μm for heating laser power just above (top) and substantially above (bottom) the bubble nucleation power threshold ($P_{532} \approx 8$ mW). b) Zoom in of transient dynamics at off-to-on transition (t=0). c) Flow dynamics in logarithmic time scale relative to application of heating laser at increasing radial distance from the antenna ($P_{532}$ = 13 mW). $\Delta t_{on}$ is the time since the heating laser has been applied in each cycle (all data shown here for 20 Hz modulation rate). d) Summary of data from 18 measurements on 6 antennas showing $r^{-3}$ radial decay of both the transient peak flow speed and the average (mean) flow speed. The measurements were performed at locations extending radially from $x_0, z_0$. Boxes extend from lower to upper quartile values, line indicates median, whiskers show the data range, and outliers are indicated as circles. Note that the Reynolds number for the transient peak is approximately $10^{-2}$.*

Optical force microscopy allows for precise velocimetry of the liquid flow with a high degree of control over the probe location, thus enabling the reconstruction of a 3D flow vector profile with high time resolution. A polystyrene (PS) microbead is kept at a variable location $r$ using a pre-calibrated holographic optical tweezers system operated at 1064 nm, far from the plasmonic resonance. By converting the displacement of the bead (Figure 1c, lower trace) to force $F(r,t)$, where we keep the velocity term from the Langevin equation to increase time-resolution by several orders-of-magnitude (see supporting information SI-1 and supporting information S4 & S5 for details), we are able to extract the speed of the liquid flow $u(r,t)$ driven by the bubble dynamics with temporal and spatial accuracy far beyond what is possible to achieve using freely diffusing tracer particles. Furthermore, this velocimetry method has the added advantage of avoiding the risk of tracer particles adhering to the bubble, which will perturb the flow profile and the bubble dynamics.

As soon as the heating laser power is increased past the bubble nucleation threshold ($P_{532} \approx 8$ mW), detected through the change in $\Delta T_{633}$, we observe strong local flows near the antenna structure (Figure 1d). No measurable flow is observed for powers below this threshold. The experimentally measured average flow profile $u_{\text{avg}}(r)$ is highly directional and points towards the nanoantenna at low angles $\phi$ relative to the glass/water interface (Figure 1e right). This flow pattern is well approximated by a Stokeslet near, and perpendicular to, a no-slip boundary (Figure 1e left). Note that the Stokeslet approximation is not specific to any particular driving mechanism but is simply the point-force solution to the Navier-Stokes equation for incompressible flow at low Reynolds number.

Figure 2 highlights the temporal and radial variation of the flow for $f_{\text{mod}}$ = 20 Hz (i.e. the duration of the on- and off-states of the heating laser cycle are 25 ms). Coinciding with bubble nucleation, signaled by an abrupt drop in $\Delta T_{633}$, we observe a strong transient flow that ceases within a few hundred microseconds after illumination turns on. The maximum flow speed in the transient, of the order 3-5 mm/s at $\phi \approx 12$ deg. and r ≈ 7 μm from the antenna centre, is approximately one order of magnitude greater than the subsequent persistent flow (Figure 2a,b).

Radial velocimetry scans (Figure 2c) reveal that both the transient and the persistent flow components decay proportional to $r^{-3}$ (Figure 2d), as expected for a Stokeslet. Importantly, the overall spatial flow profiles of the transient and the persistent components are nearly identical, i.e. directed towards the antenna at low $\phi$ and from the antenna at higher angles (see supporting information S7) and thus they serve to complement one another with for the purpose of mass transport and/or force actuation.

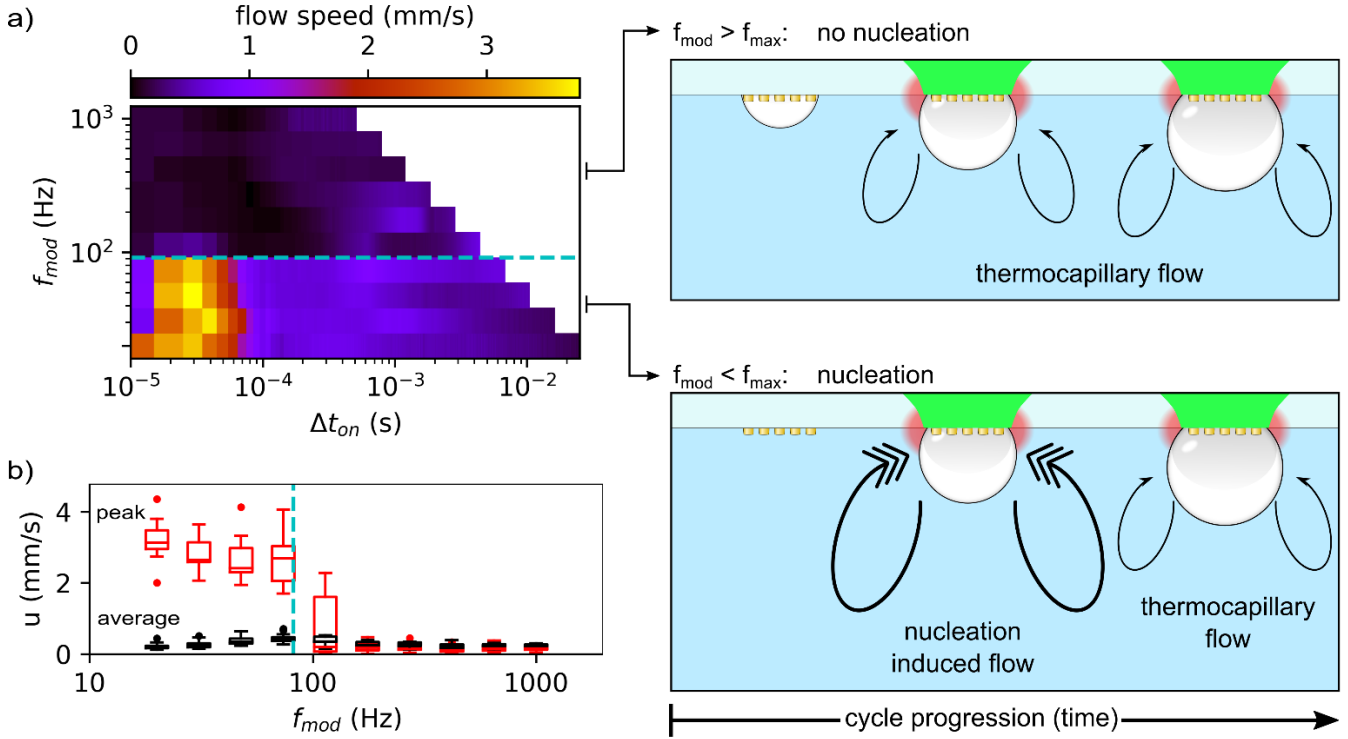

*Figure 3. The transient flow is driven by bubble nucleation. a) Flow dynamics in logarithmic time at various heat modulation frequencies. $\Delta t_{on}$ is the time since the heating laser has been applied within each heating cycle (for variable heating modulation rates as indicated by the position along the y-axis, and at a fixed probe location). The dashed blue line indicates the maximum modulation rate $f_{max}$ for the bubble to fully dissipate before the subsequent heating cycle. The panels to the right indicate the dominant process for $f_{mod} < f_{max}$ (bottom) and $f_{mod} > f_{max}$ (top). b) Combined data from 17 measurements on 5 nanoantennas showing peak transient and persistent flow speeds at $x_0$, $z_0$ versus modulation frequency. Boxes extend from lower to upper quartile values, line indicates median, whiskers show the range, and outliers are indicated as circles. All data shown here were measured for $P_{532} \approx 13$ mW at location $x_0, z_0 = 7, -1.5$ μm.*

The occurrence of the transient flow peak is intrinsically tied to the act of bubble nucleation. In Figure 3a, we show the bubble dynamics for increasing heating modulation frequency ranging from $f_{mod} = 20$ Hz to 1 kHz. It is apparent that a transition occurs at approximately $f_{max} \approx 100$ Hz, above which the transient peak is entirely absent. Importantly, the frequency region above $f_{max}$ is characterized by a measurable bubble signal $\Delta T_{633}$ throughout the off-state of the laser heating cycle (see supporting information S1c). This implies the bubble does not have time enough to fully dissipate before the start of the next heating cycle. As a result, no nucleation event occurs but instead a small bubble is already present when the heating laser turns on. The dependence of the transient peak on bubble nucleation (i.e. full bubble modulation) was consistently observed as shown in Figure 3b. In stark contrast, only a marginal effect on the persistent flow speed is observed when passing the transition frequency (see supporting information S8a).

From the Stokeslet approximation, one expects that the local flow speed is related to the primary driving force $F$ as $u \propto F/\eta$, where $\eta$ is the fluid viscosity. To investigate the mechanism of the transient peak, we measured the flow for several different glycerol-water solutions with varying $\eta$. Again, in all solutions the transient peak is absent for modulation frequencies that do not permit full bubble dissipation (Figure 4a). As seen in Figure 4b, the transient radial decay fitting coefficient $C_u$ (i.e. $u(r,\phi_0) = C_u\, r^{-3}$ for a specific $\phi_0$) shows a well-defined $\eta^{-1}$ behaviour. This implies that the primary driving force causing the transient is solution independent for the investigated water-glycerol solutions.

What is the origin of the strong flow transient? We first note that it cannot be a shockwave, which would produce a force propagating strictly outwards from the antenna at significantly shorter time scales. Secondly, the insensitivity to the host fluid (with regards to the driving force) would not be expected if the flow was purely thermocapillary in nature since the thermocapillary effect depends non-trivially on several different material properties (e.g. Henry's constant, interfacial surface tension, and their temperature derivatives, etc.). Indeed, the persistent flow component shows no clear quantitative trend between the different solutions (see supporting information S8b).

Based on these observations we hypothesize that the mechanism responsible for the transient peak is mechanical in nature and qualitatively analogous to the process utilized in thermal inkjet style actuators.[25] The spectral properties of the plasmonic antenna are nearly identical between solutions (see supporting information S9a), thus approximately the same amount of energy will have been loaded into the system in all cases. When the heating laser power is significantly above the threshold for bubble formation, which is nearly constant for all solutions (see supporting information S9b) since the boiling point increases by only $\approx 10$ K up to 50% glycerol, most of the energy associated with bubble formation will contribute to the expansion of the bubble. Thus we hypothesize that it is the rapid bubble expansion that mechanically induces the transient fluid flow. Due to the symmetry of the system, this bubble expansion can then be approximated as a point force directed perpendicular to the interface in the Stokeslet picture, which implies that it will result in the same flow profile as the subsequent persistent flow. Spatial integration of the flow profile for realistic bubble sizes and based on the Stokeslet approximation indicates that the work done on the fluid per time unit may correspond to as much as 10% of the absorbed optical power at the peak of the transient (see supporting information section SI-5).

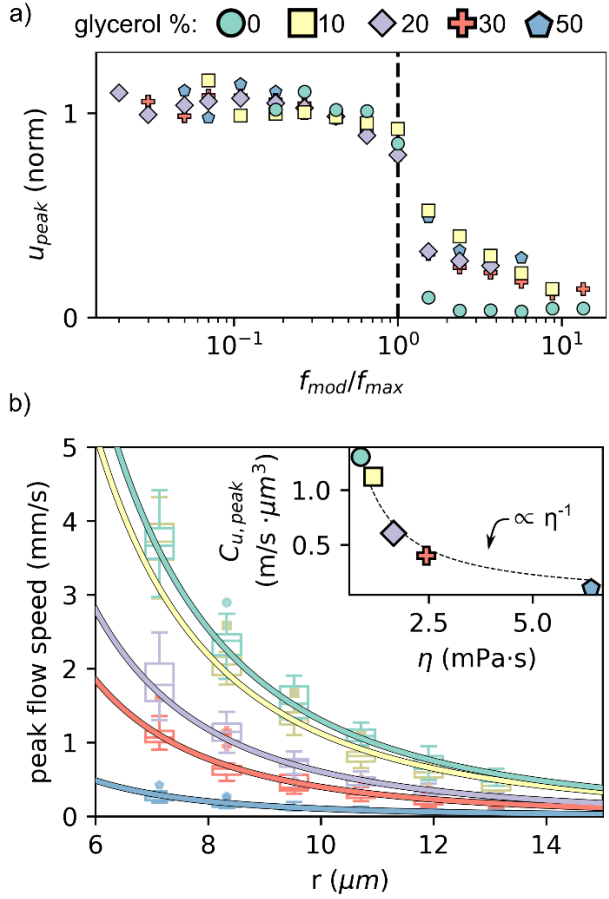

*Figure 4. Peak flow is consistent for various water-glycerol solutions. a) For all solutions there is a significant decline in the transient peak speed when full bubble dissipation cannot occur during the modulation cycle (i.e. $f_{mod} > f_{max}$). Note that the maximum modulation rate varies for each solution and all data has therefore been normalized to $f_{max}$. b) Peak flow speed radial decay for different solutions. b-inset) Radial decay fitting coefficient for each solution plotted against the solution viscosity shows a $\eta^{-1}$ dependence. Data in a) is measured at for $P_{532} \approx 13$ mW at location $x_0, z_0 = 7, -1.5$ μm (the same location as data in Figure 3) for all solutions; data in b) is measured at the same locations as the data measured in Figure 2b,c. The solution percentages indicate the glycerol content by volume. Data in a) shows the mean values for several measurements.*

The peak flow speeds/forces achievable during the transient regime of bubble formation provides an opportunity for microscale manipulation. Unlocking the full potential of photothermal surface bubbles for active manipulation in microfluidics requires further understanding and optimization of the transient flow formation. Higher modulation rates are obviously advantageous in applications where the total (cycle averaged) flow speed is the primary figure of merit. The use of smaller heating nanoantennas can enable over an order of magnitude increase in the total modulation rate by heating a smaller volume of water and thus forming smaller bubbles.[23] However, smaller bubbles would presumably entail weaker transient forces and thus the scaling of these two effects should be investigated. Directional control over the flow profile may be achieved by tailoring the antenna shape and/or the surrounding environment to influence the bubble expansion. Optimization of the peak transient force may be achieved by inducing bubble oscillations (and thus possibly inducing oscillating bubble streaming effects) during the initial bubble formation, a phenomenon that has sometimes been observed for larger microbubbles.[26,27] Other strategies for flow optimization in this system include tailoring the heat source density or utilizing solutal Marangoni effects to control the flow directionality/profile, as has been demonstrated in recent studies on larger microbubbles.[28,29]

## Conclusion

In conclusion, we have investigated the formation of flows surrounding thermoplasmonically nucleated bubbles. Specifically, the bubbles studied here are small enough such that the growth is self-limited and thus are suitable for inclusion as actuators in integrated microfluidic platforms. These bubbles have been found to exhibit two distinct flow regimes – an initial transient flow and a subsequent persistent flow. The transient flow was found to be an order of magnitude greater than the persistent flow, while both exhibit the same general flow profile making both features advantageous for a plethora of microfluidic manipulation based applications. Critically, it was found that the transient flow requires a nucleation event to occur in order to be present, thus highlighting the need for fast bubble modulation to be possible in order to fully utilize the transient phenomenon.

## Methods

### Experimental Setup

A schematic of the experimental setup is shown in S1a. The general system consists of a heavily modified upright microscope (Nikon Ti80 Eclipse) with a piezo stage (Mad City Labs, Nano-LP200) for precision alignment of the antennas and a camera for brightfield imaging used in rough alignment (Andor iXon EMCCD) and a 60×, 1.2 NA water immersion objective (Nikon CFI Plan Apo VC 60XC WI) for laser focusing and imaging. All experimental measurements were automated with LabVIEW. The optical system essentially consists of three main optical paths for the 1) heating laser, 2) bubble detection laser, and 3) optical force microscopy system.

1) The heating laser used was a 532 nm continuous wave (CW) laser (Vortran, Stradus 532-40, 532nm) which was chosen to enable strong optical absorption due to spectral overlap with the d-band electronic transition in gold. The maximum heating laser power was controlled by a voltage applied to the control unit and modulation was achieved with an acousto-optic modulator (Brimrose TEAFI10-0.4-1.0-MSD-ER). The zeroth order diffraction from the acousto-optic modulator was monitored with a silicon photodiode (Thorlabs, PDA36A-EC) while the first order diffraction was transmitted to the microscope for photothermal heating. The zeroth order photodiode signal was calibrated to provide a real-time measurement of the heating laser power reaching the sample. For all data shown here, we use square wave heating modulation with a duty-cycle of 50%.

2) Bubble detection was achieved by a lower power 632 nm CW laser (Melles Griot, HeNe) which was co-aligned with the heating laser. The transmission of this laser was collected with a condenser (Nikon CSI Plan Fluor ELWD 60XC, NA = 0.7, air) and then focused onto a silicon photodiode (Thor Labs PDA36A-EC) to monitor the transmission of the laser for both bubble detection as well as alignment with the structures. A 633 nm band pass filter and 1000 nm short pass filter was placed in front of this photodiode to prevent any leakage from the other lasers when monitoring this signal. A nonlinear response of the transmitted signal with respect to bubble size is expected, as shown in S1b (calculation details given below); however, this technique can be used to robustly detect the presence of a bubble and thus determine if a bubble is able to fully dissipate as shown S1c.

3) The optical force microscopy system consists of a 1064 nm CW laser (Cobalt, Rumba 05-01 DPSS) which was expanded to a collimated beam that slightly overfilled the back aperture of the trapping objective. The position of the trapping laser focal spot was controlled by a liquid crystal on silicon spatial light modulator (SLM) (Meadowlark Optics, 1920x1152) which was placed at a plane conjugate to the back focal plane of the objective in a 4-f configuration. The light transmitted through the trapped particle was collected by the condenser and directed toward a quadrant photodiode (First Sensor, QPD154-Q HVSD, 1064nm enhanced). The back focal plane of the condenser was imaged onto the quadrant photodiode so that the output voltage would reflect the wavevector distribution of the transmitted light. A 1064 nm bandpass filter and neutral density filters were placed in front of the quadrant photodiode to prevent oversaturation and leakage from the other laser.

**Measurement Procedure**

The measurement procedure consists of first trapping the probe particle, positioning the trapped particle at the desired location, and then aligning the heating & bubble detection lasers with the plasmonic antenna. Antenna alignment was performed via measuring the 633 nm transmission as the antenna was scanned across each axis though the 633 nm focal spot and then moving to the location of lowest transmission; this procedure was automated with LabVIEW. During the measurements, an initial "calibration" time series was obtained (typically 3 s) for the antenna aligned and probe particle in the desired position but without application of the heating laser (see S10 for comparison between experimental data and theoretical limits). After this "calibration" data was collected, the heating laser would then be modulated while data was collected on all detection channels – this consists of the "measurement" time series (typically 2 s). Data was acquired at 100 kHz with a digital acquisition device (DAQ, National Instruments, NIUSB6252). Note that both the "calibration" and "measurement" time series were recorded as one series with no break in continuity during data acquisition. After data had been acquired the system would prepare for the next measurement (e.g. change heating laser power, move probe particle, etc.). This type of serial measurement procedure for determining the flow profile is possible because of the high degree of repeatability between measurements as shown in S3. After a predetermined number of measurements had been obtained the system would automatically realign before proceeding (typically 3 measurement or $\approx$ 15 s). Although some amount of drift was present in our optical system, it was determined that drift could be limited to within 100 nm along all axes (worst case scenario) if realignment were performed at least once every 120 s. By realigning the system after substantially shorter intervals we were able to essentially eliminate any significant drift from our measurements. The standard deviation (error) of the auto-alignment procedure was measured to be 10 nm, 12.5 nm, and 50 nm along x, y, and z respectively.

Further details about the experimental procedure, including trap position calibration, trap force calibration, data analysis, and control experiments, are given in SI-1 in the supporting information.

**Thermoplasmonic Antenna: Design & Fabrication**

The nominal design of the thermoplasmonic antenna consists of 19 gold disks (height = 60 nm, radius = 50 nm) arranged in a hexagonal lattice with interparticle spacing of 150 nm (centre to centre) and an overall structure radius of approximately 350 nm. By using a collection of plasmonic disks (as opposed to a single large structure) the optical and thermal properties could be independently controlled by changing size of the individual disk and overall structure diameter, respectively (note that the disks are spaced too far apart to plasmonically couple via near-field coupling and any diffractive orders are in the ultraviolet wavelength region). The thermoplasmonic disks used here were designed such that the plasmonic resonance enabled significant absorption at 532 nm to facilitate bubble formation via photothermal heating from the heating laser while being non-resonant at the wavelength used for optical force microscopy (1064 nm ) – although in general there is essentially no interaction with the 1064 nm laser during the experiments because there is never significant geometric overlap between the trapping laser and the antenna (see SI-1 Methods & Experimental Setup - Control Experiments). The thermal properties of the antenna, dictated by the overall structure size, were designed to enable bubble nucleation at moderate power levels while simultaneously ensuring that the heating was sufficiently localized to prevent continuous bubble growth (hence ensuring high permissible modulation rates). Further details about this thermoplasmonic antenna design, and its optical/thermal properties are given in ref[23].

The disks were fabricated by electron beam lithography followed by metal deposition and lift-off. Microscope cover slides (#1.5) were washed under sonication for 5 minutes each in acetone and IPA followed by an oxygen plasma treatment (30 s at 50 W and 250 mTorr). A 150 nm thick layer of ARP 6200_13: Anisol, 1:1 resist was deposited via spin-coating and cured at 160°C. A sacrificial layer of Cr was deposited to present a reflective and conductive surface for the subsequent e-beam lithography. The nanodisks were exposed in a Raith EBPG 5200 100 kV system to a dose of ≈533 μC/cm^2. The Cr was removed by wet etching and the resist developed in n-Amylacetate for 90 s. Metals for the antennas (2 nm of Ti for adhesion and 60 nm of Au) were evaporated and then uncovered via lift-off in Remover 1165. To enhance optical properties and thermal stability, 10 minutes annealing at 150°C followed by PECVD deposition of 20 nm of SiO2 conclude the fabrication.

**Sample Preparation**

Before experiments, the coverslips containing the plasmonic antenna were rinsed with isopropanol and deionized water and dried with nitrogen. Any biological contaminants on the surface were then removed by cleaning the sample with air plasma etching (Harrick Plasma, PDC-32G) for 60 s on high. A silicone spacer (SecureSeal Spacer, 13 mm diameter, 120 μm depth) was adhered to the sample. The microwell was then filled with a dilute solution containing the probe microspheres (micro particles GmbH, diameter =1.98 μm, standard deviation =0.03 μm) in either deionized water or a deionized water + glycerol solution (Sigma-Aldrich G9012-500ML, Glycerol, 99%). All solutes are air-equilibrated before sample preparation. The sample and filled microwell was then sealed against a standard microscope slide and loaded into the experimental setup. After measurements had finished, the sample was left in an isopropanol bath until it detached from the silicone spacer and cleaned again before storage for further use.

**Bubble Detection Simulations**

Simulations of the transmitted 633 nm signal were carried out using the finite-difference time-domain method (FDTD Solutions, Lumerical). The plasmonic nanoantennas were made of gold described by a complex permittivity according to Johnson and Christy[30] and their size and layout matched the experimental dimensions: hexagonal pattern with a lattice of 150 nm composed of 19 nanodisks with vertical side walls 60 nm in thickness, 5 nm top and bottom edge rounding, and diameter of 95 nm (slightly smaller than the nominal design radius). They were placed on a substrate with a refractive index of 1.5 and covered by a conformal 20 nm thick dielectric layer with index of 1.5. The medium on top of the structure is water with n = 1.33. Illumination of the structures was from the substrate side with a broad-spectrum plane wave to match the simulated spectrum to the experimental one and a Gaussian beam with a waist radius of 800 nm to model transmission changes at 633 nm induced by bubble growth. Permittivity changes with temperature were neglected. The forward scattered signal was collected over a large area (30 µm × 30 µm, 2.4 µm from the substrate) to perform far-field propagation to account for the collection efficiency of the objective. The bubble was modelled as an air-filled sphere with a variable radius (100-1000 nm) and a variable contact angle (30, 60, 90, 120 degrees). The mesh size around the nanodisks in a volume 800 nm × 700 nm × 130 nm was set to 2 nm, followed by a gradual mesh increase to 8 nm and then 12 nm. Symmetric and antisymmetric boundary conditions were also used.

The transmitted signal for all studied contact angles has a similar dependence with bubble size. For small bubbles with radii less than approximately 400 nm the transmitted signal initially increases. Evidence of this transmission increase for small bubbles can be inferred from S1c where it can be seen that as the bubble dissipates, the transmission briefly increases before bubble dissipation is complete. Once the bubble size exceeds the lateral dimensions of the nanostructure, the signal begins to decrease more quickly than the initial increase. This decrease is non-monotonic and Fabry-Perot-like oscillations are present. In general, the slope of this decrease in transmitted signal intensity is proportional to the contact angle, which determines the overall bubble size. The slope of the signal dependence is the largest for 30 deg contact angle and decreases ca. 5-fold for the largest studied angle of 120 deg.

**Author contributions:** S.J. and M.K. conceived the study. S.J., A.S., and H.R.D. designed the experimental setup. S.J. performed the measurements and analysed the results. D.A. fabricated the samples. T.J.A. performed optical simulations. S.J. and M.K. interpreted the data and wrote the paper with input from all other authors.

**Acknowledgements:** This work was funded by the Swedish Research Council and Chalmers Excellence Initiative Nano.

**Competing Interests:** The authors declare no competing interests.

**Supporting Information:** The Supporting Information is available free of charge online. Further details about experimental setup and data analysis technique; control experiments; xz-flow profile for transient peak and persistent flow; modulation rate effect on persistent flow in aqueous glycerol solutions; darkfield spectra of nanoantennas and bubble threshold in aqueous glycerol solutions; and details about energy dissipation calculation.

# Supporting Information for: Strong Transient Flows Generated by Thermoplasmonic Bubble Nucleation

Steven Jones, Daniel Andrén, Tomasz J. Antosiewicz, Alexander Stilgoe, Halina Rubinsztein-Dunlop, & Mikael Käll

## SI-1 Methods & Experimental Setup

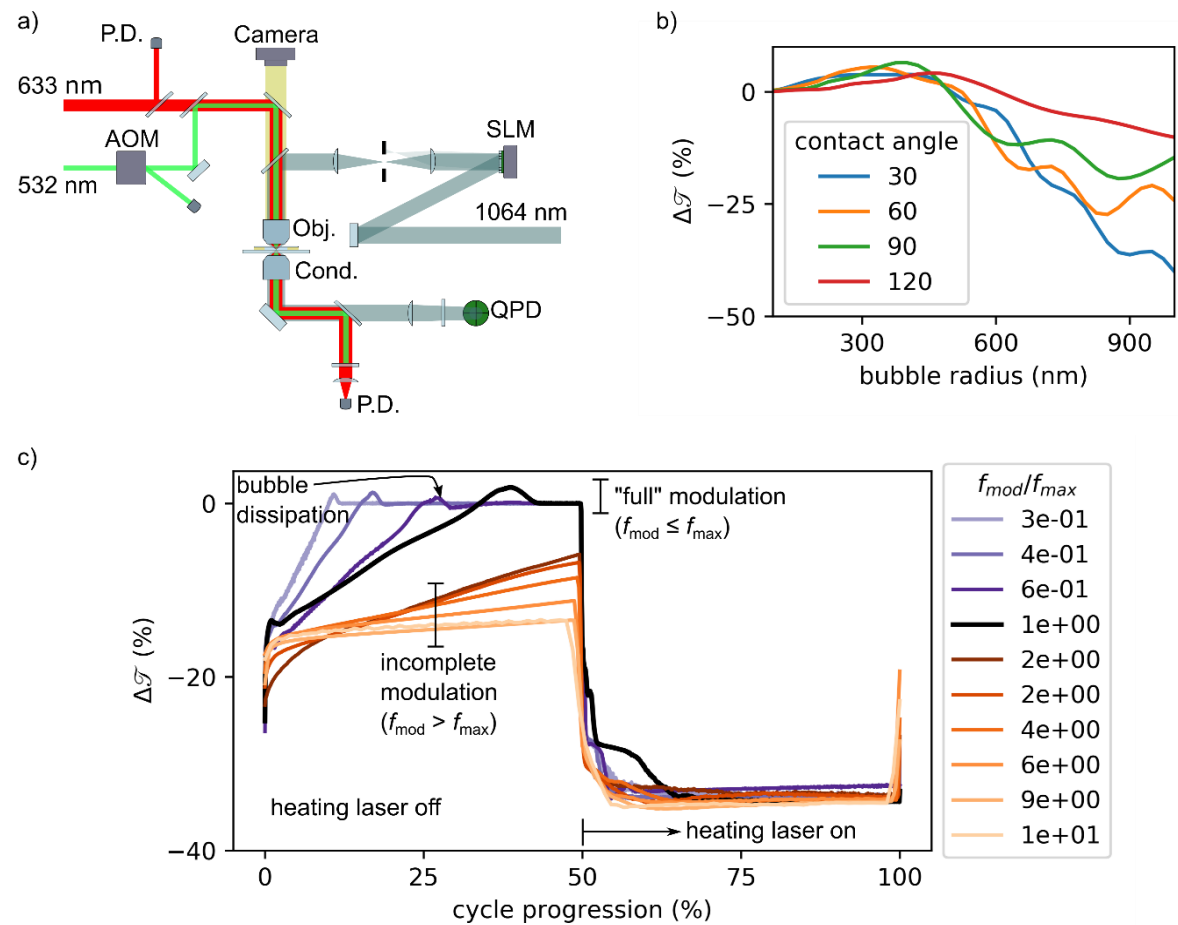

S1 a) Experimental setup. Terms: P.D. – photodiode; AOM – acousto-optic modulator; Obj. – imaging/trapping objective; Cond. – condenser objective; SLM – liquid crystal on silicon spatial light modulator; QPD – quadrant photodiode. b) Simulated bubble detection signal in water based on FDTD simulations for varying contact angles (measured between glass and bubble surface through the water). c) Experimental bubble signal response for various modulation frequencies in water. Each modulation cycle is normalized to percent of full cycle to highlight the different modulation responses on a single plot. Approximate values for the normalized modulation frequency (with respect to the maximum possible modulation frequency for "full" modulation) are shown on the right-hand side. The purple traces show $f_{mod} < f_{max}$ while the orange traces show $f_{mod} > f_{max}$, the black trace shows $f_{mod} \approx f_{max}$. For modulation frequencies above the maximum value the bubble can clearly be seen to never fully dissipate during the cycle as evidenced by the significant change in transmission still present when the heating laser is applied at cycle progression = 50 %.

## Optical Force Microscopy – Position Calibration

The position of the optical trap relative to the plasmonic antenna was carefully calibrated prior to experimental measurements. This calibration was performed by performing linear ramps of the phase gradient corresponding to each geometric axis (linear phase ramps in x and y, and phase curvature in z). For each phase condition the focal spot was imaged via reflection from the glass-water interface with the camera (which had previously been calibrated for the image pixel size) to determine its location within the xy plane. The piezo stage independently determined the depth of the trap along z. This resulted in a 3D array of phase conditions corresponding to a 3D array of focal spot locations. Each axis was then fit with 2$^{nd}$ order polynomials to get a function describing the geometric focal spot position in terms of the SLM phase profile. Additional 2$^{nd}$ degree polynomials were also included to correct for any slight rotational offset between the camera and the SLM as well as to account for the varying magnification with z-position. The resulting calibration was then tested by mapping out a $x: \pm 15$ μm $\times y: \pm 15$ μm $\times z: +5, -15$ μm region surrounding the nominal antenna location. At each position, the focal spot location was measured and compared against the "desired" position to determine the accuracy of the SLM calibration. The calibration/testing procedure was automated with LabVIEW and resulted in trap position error residuals with standard deviations: $\sigma_{\varepsilon,x} = 63$ nm, $\sigma_{\varepsilon,y} = 52$ nm, and $\sigma_{\varepsilon,z} = 69$ nm.

## Optical Force Microscopy – Trap Calibration

The stiffness of the optical trap and conversion factor of the QPD for each axis was acquired *via* power spectrum analysis similar to the procedure outlined in ref[1]. The power spectral density (PSD) fitting procedure was tested for the case of both a simple Lorentzian profile as well as for the full hydrodynamic model resulting in nominally identical results – see S2. The Lorentzian model is sufficient in our case for two reasons: first, our detection bandwidth is not high enough to measure the frequency range where hydrodynamic effects become dominant in bulk (*i.e.* $\geq 1$ MHz) and second, the relatively close proximity of our probe particle to the surface causes hydrodynamic effects to be suppressed until still higher frequencies. Therefore in all cases the low frequency approximation for the drag coefficient is sufficient. In order to account for boundary effects on the low frequency drag coefficient Faxen's approximation was used for parallel motion and the infinite series derived by Brenner was used for perpendicular motion, *i.e.*:[2,3]

$$\gamma_\parallel = \gamma_0 \left[ 1 - \frac{9}{8}\frac{a}{2h} + \left(\frac{a}{2h}\right)^3 - \frac{45}{16}\left(\frac{a}{2h}\right)^4 - 2\left(\frac{a}{2h}\right)^5 \right]^{-1}$$

and

$$\gamma_\perp = \gamma_0 \left[ \frac{4}{3}\sinh(\beta) \sum_{n=1}^{\infty} \left\{ \frac{n(n+1)}{(2n-1)(2n+3)} \times \left( \frac{2\sinh(\beta(2n+1)) + (2n+1)\sinh(2\beta)}{4\sinh^2\left(\beta\left(n+\frac{1}{2}\right)\right) - (2n+1)^2 \sinh^2(\beta)} - 1 \right) \right\} \right]$$

where $\gamma_0 = 6\pi\eta a$ is Stokes drag coefficient, $a$ is the microsphere radius, $h$ is the displacement of the bead from the interface (centre-to-interface), and $\beta = \cosh^{-1}\left(\frac{h}{a}\right)$. The perpendicular drag coefficient was truncated after 80 terms.

This method of optical force microscopy calibration assumes that the potential is harmonic and the wavevector changes linearly with displacement. In order to confirm this assumption we checked the linear range by translating a microparticle stuck to the glass slide through the trapping laser focal spot to measure the QPD response as a function of particle displacement. We then compared all experimentally measured displacements and confirmer that we were well within the linear range for our system. This calibration procedure results in fitted values corresponding to the conversion factor $\beta$ enabling the QPD voltage measurements to be converted to particle displacement values and the corner frequency $f_c$ which can be used to determine the trap stiffness along each axis according to $\kappa_i = 2\pi \gamma_i f_{c,i}$ where $\gamma_i$ is the low frequency drag coefficient (with boundary effects) along the corresponding axis.

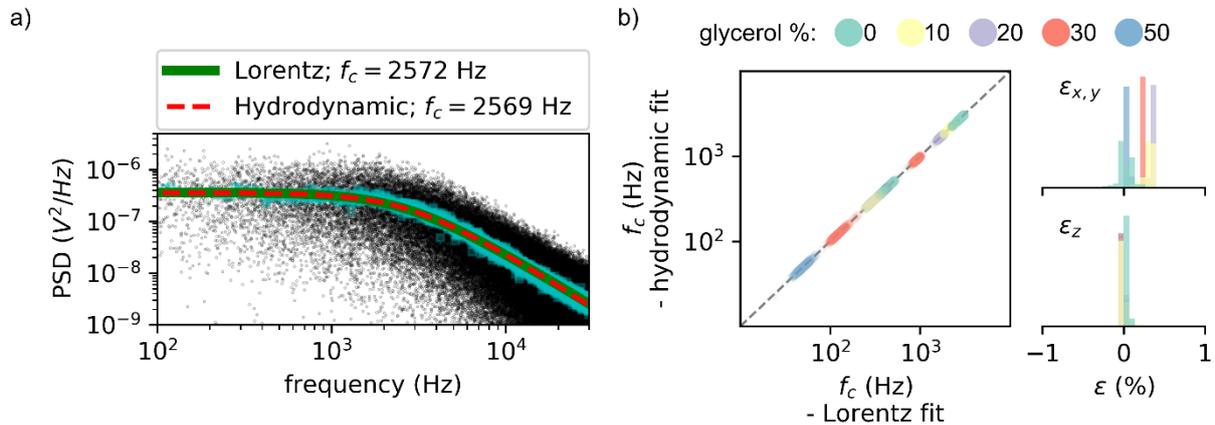

*S2 Hydrodynamic and Lorentz PSD fitting comparison. a) Demonstration of the PSD calculated by periodogram (black) and Welch's method (blue) with the corresponding numerical fits to the data via the Lorentz approximation (green) and by using the full hydrodynamic model (red). b) Comparison of extracted fit parameters obtained via the two fitting methods from several measurements in solutions with varying glycerol content. The main figure shows the extracted corner frequency obtained via the hydrodynamic model vs the same parameter extracted from fitting the PSD to a Lorentzian. The dashed grey line indicates the line of perfect agreement. The subplots on the right show the relative difference between the two methods in the x/y directions (upper subplot), and in the z-direction (lower subplot).*

Data Analysis

Five channels were measured during data acquisition corresponding to: heating laser power, bubble-detection laser transmitted intensity, and the QPD signal corresponding to particle displacements along each geometric axis. The heating laser power was converted to mW from volts by using previous calibration values obtained *via* a power meter. The bubble-detection laser transmission signal is normalized to the mean value during the "calibration" phase and only reported as relative changes in transmitted intensity. In order to obtain position, force, or effective flow measurements from the QPD signal, it first requires calibration *via* the method described above (see Optical Force Microscopy – Trap Calibration). In addition to the calibration of the trap, the phase instability inherent to all SLMs must also be removed. In our case the SLM produced a relatively stable phase output with variations less than about 2%. Regardless, due to the tight confinement of our particle within the trap, these effects still need to be removed from the resulting particle trajectory time series. The phase instability manifests itself as oscillations in the 1st order diffracted intensity as well as very slight oscillations in trap location along one of the SLM axes. These oscillations occur at a fundamental frequency as well as all harmonics with decreasing amplitude at higher orders. In order to remove the effect, the QPD data was analysed at each harmonic to determine if the oscillation was observable above the local noise level, and if so a high quality factor Fourier notch filter was digitally applied to remove the phase oscillation while leaving the power spectrum, and time series, otherwise undisturbed.

Next, the particle trajectory along each axis was averaged over all measured cycles with time $t = 0$ corresponding to the instant the heating laser is turned on. For measurements where the acquisition rate is not an integer multiple of the modulation frequency the cycle average data was binned/averaged such that the output "cycle averaged" time series would have data points spaced out by $\delta t = 1/f_s$ (where $f_s$ is the sampling frequency). Figure S3 illustrates the repeatability of these measurements both within a single measurement series and between different measurements under nominally identical conditions.

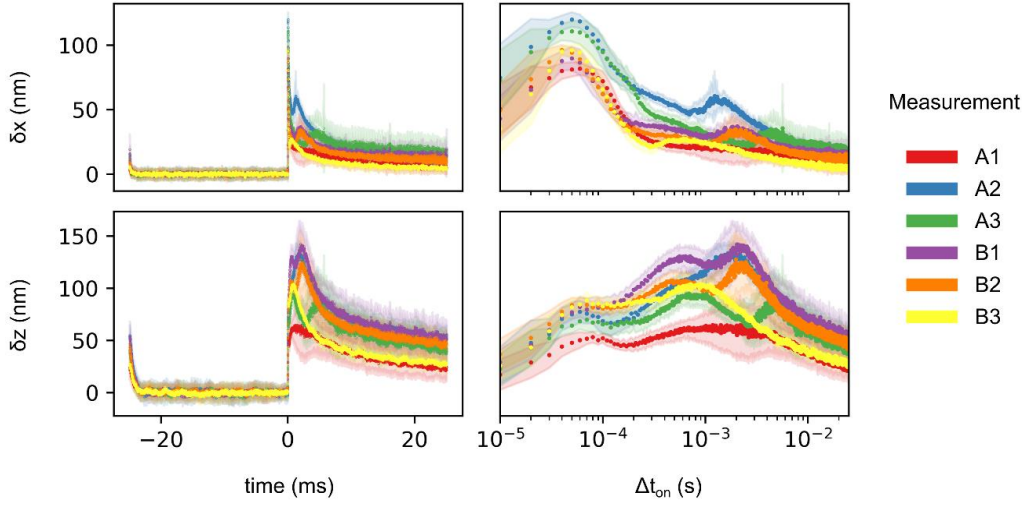

*S3 Measurement repeatability. Shown here are the probe particle position time series from six measurements on two different, nominally identical, antennas (A & B). Top row shows the probe particle displacement along the x-axis while the bottom row shows the displacement along the z-axis (optical axis). Note that these measurements correspond to the same measurement conditions as Figure 3 of the main text ($P_{532} \approx 13\ mW$ at location $x_0, z_0 = 7, -1.5\ \mu m$) and the displacement along the y-axis is negligible at all times. The markers (sphere) indicate the cycle averaged probe particle position at each time. The shaded region illustrates $\pm$ one standard deviation from the cycle averaged position to illustrate the consistency between cycles within the same data series. The different colors indicate different measurement series under nominally identical conditions to illustrate the repeatability between data sets.*

Typically in optical force microscopy measurements, it is assumed that the maximum time resolution is given by $1/f_c$ since at shorter time scales the energy is stored in the drag force opposing particle motion as opposed to in the spring force of the optical trap. However, because we are using nearly perfect spherical probe particles and therefore we know the drag force acting on the particle if we know the trajectory, we can utilize this to obtain force measurements within our system that is limited only by our DAQ (as well as the bandwidth of optical and electronic components). In other words, we monitor the particle trajectory and then extract the external force acting on the particle as

$$F_{ex,i} = \kappa_i x_i + \gamma_i \frac{dx_i}{dt}$$

This results in force measurements with a time resolution equal to the sampling frequency, albeit at the expense of increased noise since the velocity autocorrelation for Brownian motion can only be resolved as a delta function at our sampling rate. This increased noise can then be filtered out (*i.e.* with a median filter), thereby resulting in a signal with good time resolution and signal to noise ratio.

To test this procedure we performed a control experiment were a trapped particle was pushed by a defocused secondary laser (*i.e.* no gradient force contribution from the pushing laser). The pushing laser was modulated with a square wave temporal profile. Since the pushing force is mediated by photons it is therefore proportional to the pushing laser intensity, and the modulation rise time is significantly faster than the temporal resolution of our sampling frequency – thus it should be a perfect square wave. As a result, the procedure above should be able to recover this pushing force as a square wave regardless of the corner frequency of the trapping laser. In S4 we demonstrate that we are indeed able to recover this square wave force profile, therefore validating this methodology for transient force analysis in optical tweezers.

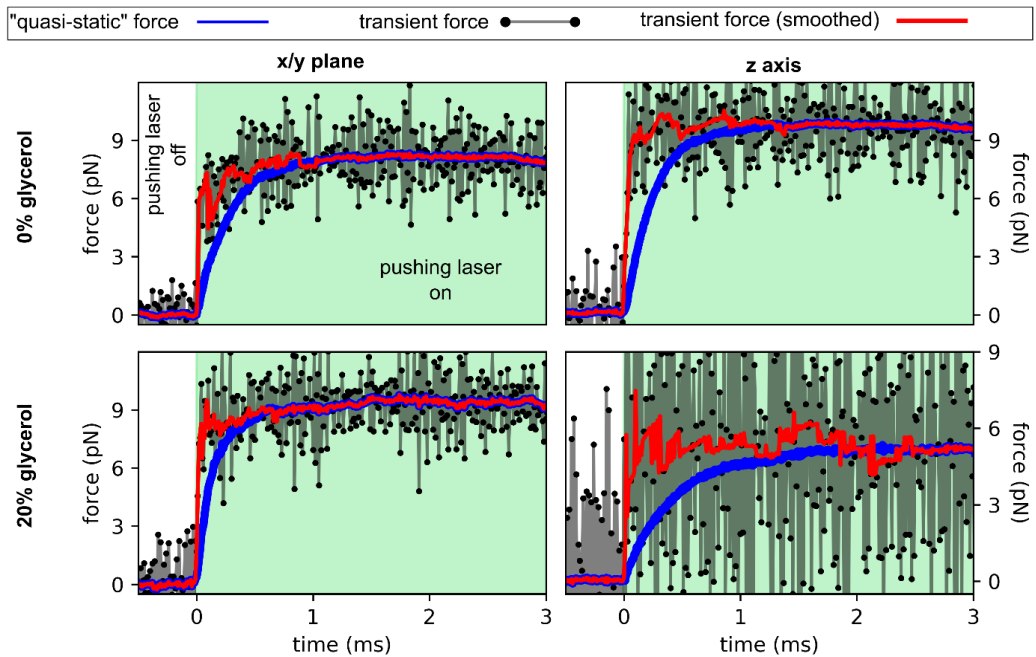

*S4 Transient force measurements with optical tweezers. Normally, the force measured by optical tweezers assumes steady-state conditions and the force on the trapped particle is simply given by the particle displacement multiplied by the stiffness ("quasi-static" force estimation, blue). However, by including the velocity term from the Langevin equation, transient forces can be recovered (transient force, black) albeit at the expense of increased noise in the measurements. These transient force measurements can be smoothed to reduce the white noise amplification (transient force (smoothed), red) which results in a minimal decrease in the time resolution in exchange for a significant reduction in the noise amplification. Data shown here is for a test case where a trapped microsphere was pushed by a secondary laser that was turned on/off in a step-like manner. Data is shown for forces in the x/y plane (left) and along the z axis (right), and for measurements performed in both pure water (top) and 20% glycerol solution (bottom).*

In data analysis for the particle displacement due to thermoplasmonically induced bubble flow this analysis methodology is utilized to recover the transient force present after bubble nucleation regardless of the trap corner frequency. A demonstration of this analysis is shown in S where it is shown that for a measurement with high corner frequency (*i.e.* S – top left) the transient features are observable but under-estimated at short time scales. For other measurements with a lower corner frequency (*i.e.* S top right and bottom row) the transient peak is still observable but is "smeared out" in time; however, after application of this analysis procedure the transient peak can once again be recovered. The recovery of transient forces in optical tweezers with time resolution greater than $1/f_c$ can be easily achieved simply by including additional terms in the Lorentzian equation of motion – specifically the velocity term. After recovery of the transient forces, we can then convert these forces into an effective local flow speed, *i.e.*

$$u_{\text{eff},i} = \frac{F_{\text{ex},i}}{\gamma_i}$$

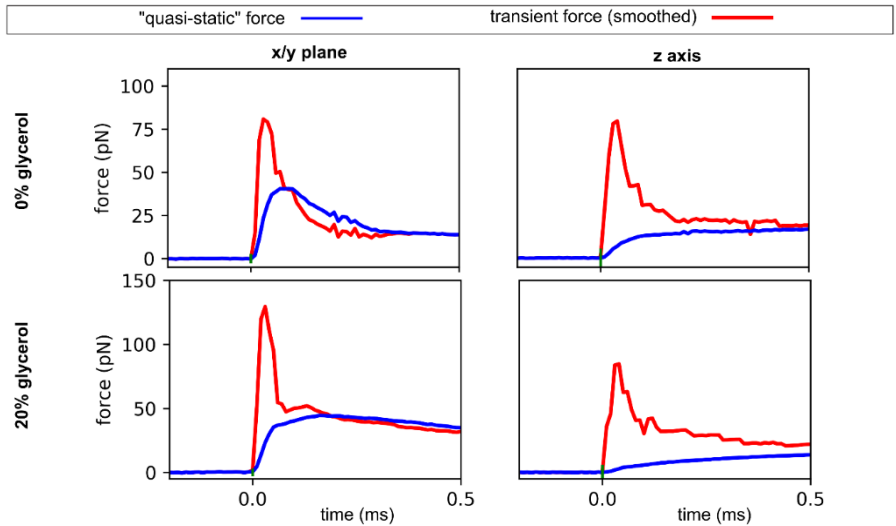

*S5 Transient analysis of bubble induced force on trapped particles. Shown here is the same analysis as demonstrated in the "square force test" above but applied to experimental data on the bubble induced force on a trapped microsphere. Although the nominal "quasi-static" method is suitable for longer time scales, it either underestimates, or entirely misses, transient forces that occur before the trap can reach steady-state conditions. By utilizing both the displacement of the particle as well as the particle velocity in calculating the force we can properly observe these transient effects.*

## Control Experiments

We have also performed several control experiments to demonstrate that none of the effects observed are the result of either probe particle interaction with the heating/bubble-detection lasers or due to interaction between the trapping laser and bubble resulting in changes in the transmitted wavevectors. In all cases we can confirm that within the region measured these effects are not present and therefore the results observed are only due to the effect of bubble nucleation and the resulting fluid flow acting on the trapped particle – see S6.

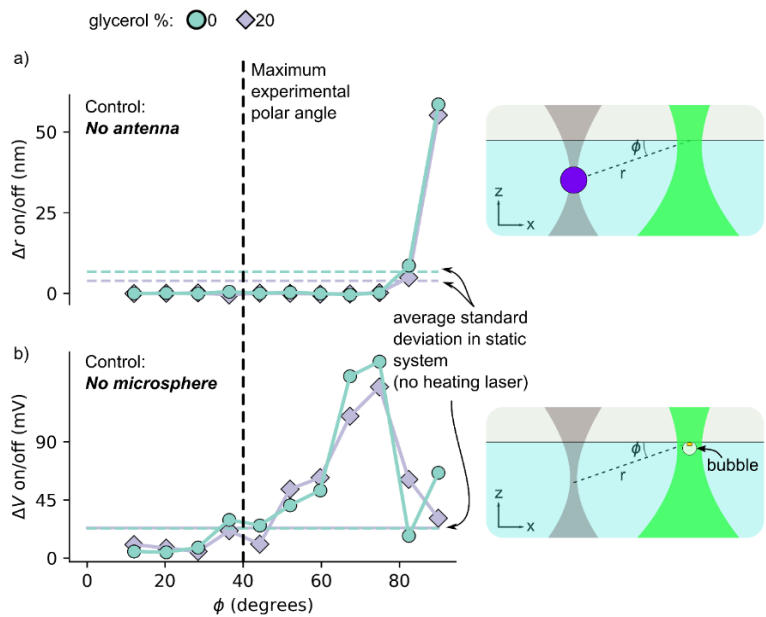

*S6 Control experiments. a) Control experiment measuring the effect of the heating laser modulation on the probe particle position within the trap. Here no antenna or bubble is included. b) Control experiment measuring the effect of the bubble on the trapping laser transmission (k-vector) in the absence of a probe particle. Dashed lines in a) and b) show the standard deviation of the respective measurement with no heating laser applied. In both control experiments there is negligible effect on the measurements for the range of polar angles investigated ($\phi_{max} \approx 40°$). All data shown here is for a nominal probe particle displacement of $r = 7\ \mu m$.*

## SI-2 Transient and Persistent Flow Profile

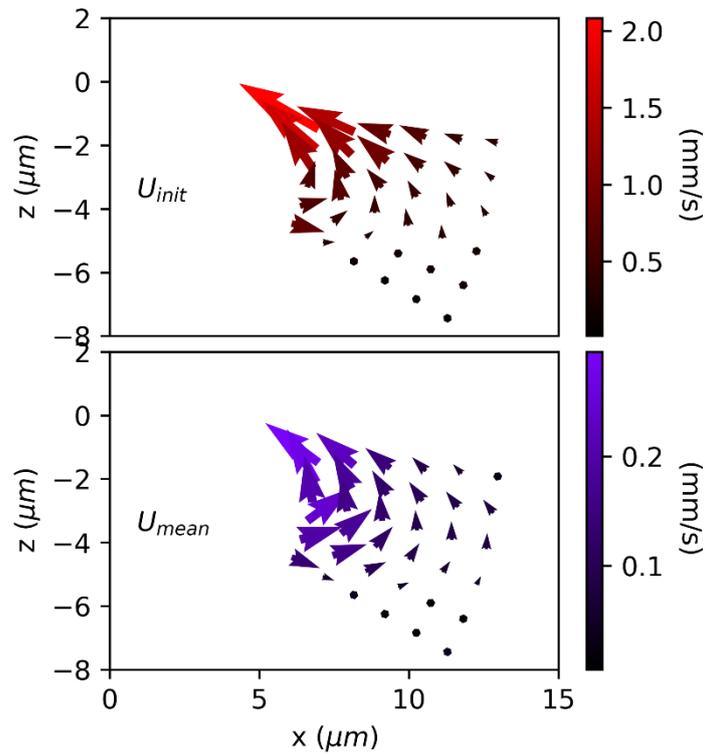

*S7 xz-plane flow profile comparison between the transient flow peak (top) and cycle averaged flow (bottom). The transient flow speed is time averaged over the first 100 μs of the cycle, while the cycle averaged flow is averaged over the entire duration of the "on" state.*

## SI-3 Persistent Flow Speed at Different Modulation Frequencies

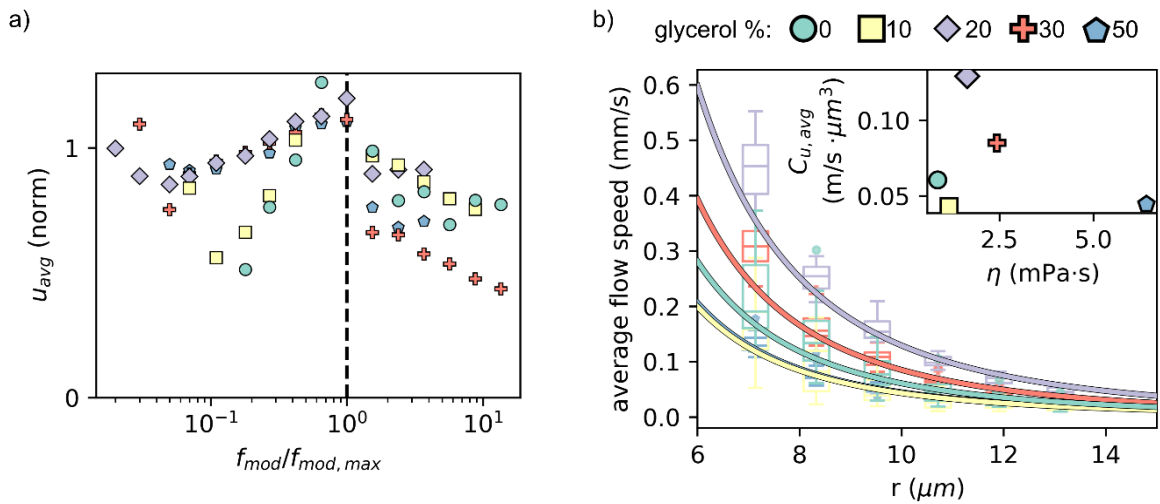

*S8 a) Average flow speed modulation frequency dependence (relative to maximum modulation rate). b) Comparison of the persistent flow speed radial decay for aqueous glycerol solutions. Inset shows the fitting coefficient for each solution.*

# SI-4 Bubble Formation in Different Solutions

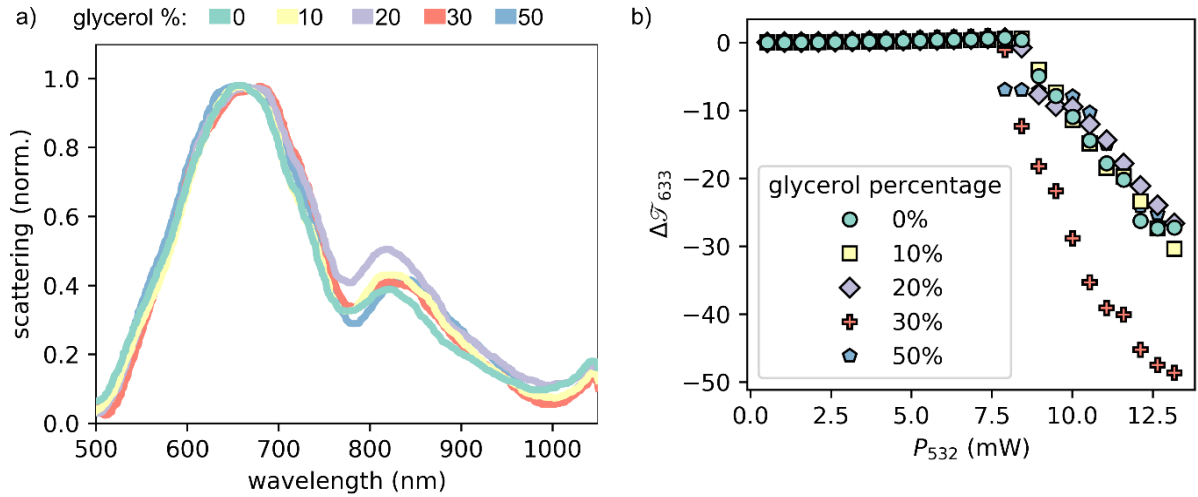

S9 a) Experimental darkfield spectra of plasmonic nanoantennas. Darkfield spectra show the same nominal optical properties regardless of solution. b) Bubble generation for various water-glycerol solutions as indicated by bubble detection signal. For all solutions bubble nucleation occurs at approximately the same heating laser power.

# SI-5 Estimating Energy Dissipation

Under the Stokeslet approximation[4] the flow speed resulting from a force singularity near an infinite no-slip interface, with the force vector normal to the interface, is given by:

$$u_x = \frac{F}{8\pi\eta}\left[\frac{x(z-h)}{r^3} + \frac{x(z+h)}{R^3} + 6h^2\frac{x(z+h)}{R^5} + 2hx\left(\frac{1}{R^3} - \frac{3(z+h)^2}{R^7}\right)\right]$$

$$u_z = \frac{F}{8\pi\eta}\left[\left(\frac{1}{r} + \frac{(z-h)^2}{r^3}\right) + \left(\frac{1}{R} + \frac{(z+h)^2}{R^3}\right) - 2h^2\left(\frac{1}{R^3} - \frac{3(z+h)^2}{R^5}\right) + 2h\left(\frac{(z+h)}{R^3} - \frac{3(z+h)^3}{R^5}\right)\right]$$

where $h$ is the displacement of the force singularity from the interface and $r, R$ is the distance to the force singularity and the corresponding image singularities respectively, *i.e.* $r = \sqrt{x^2 + y^2 + (z-h)^2}$ and $R = \sqrt{x^2 + y^2 + (z+h)^2}$. Note that the flow is rotationally symmetric about the axis normal to the interface which intersects the force singularity. The energy dissipation to heat resulting from viscous stresses in non-turbulent flow is given by:[5]

$$Q = -\frac{1}{2}\eta \int \left(\frac{\partial u_i}{\partial x_j} + \frac{\partial u_j}{\partial x_i}\right)^2 dV$$

Thus, by numerically solving the above expression for the flow field induced by a Stokeslet an estimation for the total power dissipated by the flow can be obtained by assuming quasi steady state conditions. Note that the magnitude of the force component, $F$, and the Stokeslet displacement, $h$, were obtained by numerically fitting the experimental data to the Stokeslet flow field resulting in values of 120 nN and 1.35 μm respectively. Once an estimate for the heat dissipated by viscous stresses has been obtained, one can compare this with the power absorbed by the nanoantenna to estimate the conversion efficiency. Using estimates for the absorption cross section ($\approx 21 \times 10^4$ nm$^2$) of our antenna and the incident intensity ($\approx 1.5 \times 10^9$ W/m$^2$) and numerical evaluation of the energy dissipation of the fluid flow field, the conversion efficiency at the transient peak was estimated to be approximately 10% with the remaining energy being primarily dissipated as heat *via* conduction from the antenna.

## SI-6 Probe Particle Equilibrium Behaviour in Static Optical Trap.

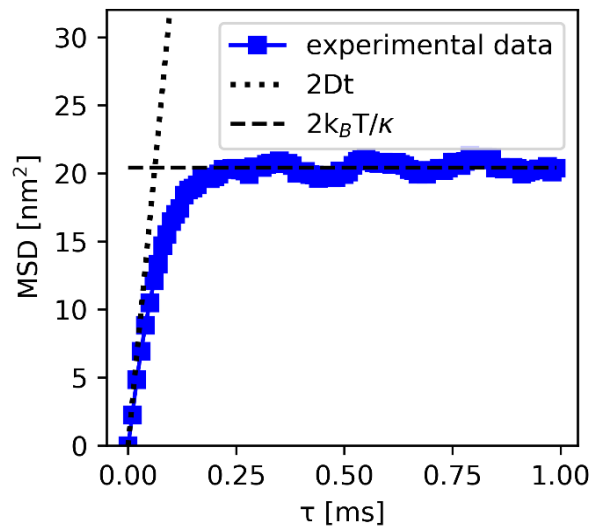

*S10 Experimental mean squared displacement (blue squares) compared to the theoretical limits for a Brownian particle in a harmonic potential at short time scales (dotted line) and the asymptotic limit (dashed line).*